\documentclass[letterpaper,12pt]{article}
\usepackage{graphicx}
\usepackage{amsmath,amsfonts}

\def\tu{{\bar u}}
\def\tv{{\bar v}}
\def\tx{{\bar x}}
\def\tz{{\bar z}}
\def\tr{{\bar r}}
\def\Tt{{\bar t}}
\def\C{{\cal C}}
\def\M{{\cal M}}

\def\d{{\mathrm d}}

\title{ Horizon Formation\\ in High-Energy
 Particles Collision }
\author{O.~I.~Vasilenko\thanks{E-mail address:
\texttt{vasilenko@depni.sinp.msu.ru}}
\\[6pt]
\emph{\small Department of Physics, M.~V.~Lomonosov Moscow State University,}\\
\emph{\small  Moscow 119992, Vorob'evy Gory, Russia}}
\date{}

\begin{document}

\maketitle \abstract{We investigate a classical formation of a
trapped surface in 4-dimensional flat space-time in a process of a
non-head-on collision of two high-energy particles which are
treated as Aichelburg-Sexl shock waves. From the condition of the
horizon volume local maximality an equation for the trapped
surface is deduced. Using a known solution on the shocks we find a
time-dependent solution describing the trapped surface between the
shocks. We analyze the horizon appearance and evolution. Obtained
results may describe qualitatively the horizon formation in higher
dimensional space-time.}

\section{Introduction}\label{introduction}

There is a significant interest to  the  processes of a black hole
production in ultra-relativistic particle collisions. Expectations
exist that on small distances space-time has more than four
dimensions, the Planck scale lowers to ${\cal O}(TeV)$ and black
holes can be produced at future accelerators
\cite{{Giddings:0709}, Kanti:0802}.

Eardley and Giddings in \cite{Eardley/Giddings:0201} proposed a
classical description of a $D$-dimensional black holes production
in a high-energy collision of two particles which are treated as
Aichelburg-Sexl shock waves. They used a trapped surface method
and constructed an analytical solution for a non-head-on collision
in $D=4$. Yoshino and Nambu in \cite{Yoshino/Nambu:0204} tried to
use this method to describe a horizon dynamic in a head-on
collision but made a mistake in border conditions. The correct
results were obtained in \cite{Vasilenko:0305}.

In this work we use the $(D=4)$-solution of
\cite{Eardley/Giddings:0201} as border conditions on the shocks to
obtain an analytical solution for the trapped surface between the
shocks. This time-depended solution describes a formation and
dynamic of the horizon at non-zero impact parameter.

The paper is organized as follows. In Section~\ref{Shock waves
metric}, we describe the Aichelburg-Sexl metric. In
Section~\ref{Outer trapped surface}, we reformulate the
$(D=4)$-solution (border conditions) of
\cite{Eardley/Giddings:0201} on the shocks in a form convenient
for further calculations. In Section~\ref{Inner trapped surface},
we use an extremum volume method to obtain an equation for the
inner trapped surface. We find a solution satisfying the border
conditions on the shocks. In Section~\ref{Trap surface formation},
we analyze this solution and the appearance and further evolution
of the trapped surfaces.

\section{Shock waves metric}\label{Shock waves metric}
We use a Minkowski coordinate system
$(\tu=\Tt-\tz,\tv=\Tt+\tz,\tx^i)$, $i=1,2$. The particles are
moving along $\tz$ axis  in the opposite directions with an impact
parameter $b$ and  transverse coordinates ($\tx^1=\pm b/2,
\tx^2=0$).

The gravitational solution for a particle of total energy $\mu$
moving in the $+z$ direction with $\tx^{1,2}=0$ is the
Aichelburg-Sexl metric \cite{Aichelburg/Sexl:71,Dray/Hooft:85}
\begin{equation}
\label{Aichelburg-Sexl} \d{}s^2 = -\d\tu \d\tv + \d\tx^{i2} +
\Phi(\tx^i) \delta(\tu) \d\tu^2\ .
\end{equation}
Function $\Phi= -2a\ln(\tr)$ depends only on the transverse radius
$\tr= \sqrt{\tx^i\tx_i}$. Here $a=4G\mu$ and $G$ is the
gravitational constant.

It is possible to remove singularity in the metric
\eqref{Aichelburg-Sexl} by introducing new coordinates ($u,v,x^i$)
defined by
\begin{equation}
\label{Coordinates} \tu = u\ ,\ \ \tv = v+\Phi\theta(u) + \frac{u
\theta(u) (\nabla\Phi)^2}{4}\ ,\ \ \tx^i= x^i + \frac{u}{2}
\nabla_i \Phi\theta(u)\ .
\end{equation}
(here $\theta$ is the Heaviside step function). In these
coordinates,  geodesics and their tangents are continuous across
the shock plane at $u=0$.

Metric \eqref{Aichelburg-Sexl} is flat everywhere except  the null
plane $\tu=0$ of the shock wave. So, in order to obtain a two
shock waves metric for time $\Tt<0$ preceding the collision  we
can combine it with another similar metric corresponding to the
particle moving along $\tv=0$ in the $-z$ direction by  matching
together the regions of flat space which precede each of two
waves.

\section{Outer trapped surfaces}
\label{Outer trapped surface}
 We shall consider the following slice of space-time:
\begin{align}
&{\rm region~R_+:} \quad t=z,\qquad t\leq T\ ,\notag\label{Space-TimeSlice}\\
&{\rm region~R_I:} \quad t=T,\quad T\leq z \leq -T\ ,\\
&{\rm region~R_-:} \quad t=-z,\quad t\leq T\ .\notag
\end{align}
Here $T\leq 0$ and the collision of the shocks takes place at
($T=0,\ z=0$).

Define outer trapped surfaces $\M_+$ and $\M_-$ in the
regions~$R_+$ and $R_-$ by expressions
\begin{align}
\label{OuterTrapExpr} u=+0\ ,\quad  v= -\Psi_+(x^1,x^2)\ \qquad
{\rm and}\qquad v=+0\ ,\quad u&= -\Psi_-(x^1,x^2)
\end{align}
correspondingly.

At the moment of the collision the trapped surfaces $\M_{\pm}$
must combine into a unified continuous and smooth trapped surface.
This demand is sufficient for the determination of the functions
$\Psi_{\pm}$. They were found in \cite{Eardley/Giddings:0201} and
may be written as
\begin{equation}
\label{Functions-Psi} \Psi_{\pm}=-a\ln\left[\frac{e^{2k(\pm
x^1-b/2)}-2e^{k(\pm x^1-b/2)}\cos(kx^2)+1}{1-e^{-kb}}\right]
\end{equation}
where $k$ is determined by
\begin{equation}%
\label{Function-k} e^{kb}-1=(ka)^2\ .
\end{equation}

The trapped surfaces $\M_{\pm}$ intersect  borders $(z=\pm T)$ of
the region~$R_I$  in closed contours $\C_{\pm}$ defined by
expressions
\begin{equation}\label{Contours-C}
\cosh[k(x^1\mp x_b)]=e^{k(b/2-x_b)}\cos(kx^2),\quad
1-e^{k(x_b-b/2)}=\left(1-e^{-kb}\right)e^{2T/a}\ .
\end{equation}

\section{Inner trapped surface}
\label{Inner trapped surface}
According to the slice \eqref{Space-TimeSlice}, we define a
trapped surface $\M_I$\ in the region~$R_I$ as
\begin{equation}
\label{InnerTrapExpr}
 t=T=const\ , \qquad z= f(x^1,x^2)\ .
\end{equation}
A null normal $N$ to this surface is given by
\begin{equation}
\label{InnerTrapNullNormals}
 N= \left[N^t,\ N^z,\ N^{x^1},\ N^{x^2}\right]=
\left[1,  \frac{1}{\sqrt{1+|\vec\nabla f|^2}},\ -\frac{
f_{x^1}}{\sqrt{1+|\vec\nabla f|^2}},\
 -\frac{f_{x^2}}{\sqrt{1+|\vec\nabla f|^2}}
\right]
\end{equation}
where $|\vec\nabla f|=\sqrt{f_{x^1}^2+f_{x^2}^2}$.

A null geodesic which normally crosses the trapped surface
\eqref{InnerTrapExpr} at ($T, x^1_0, x^2_0, z_0=f(x^1_0,
x^2_0)\equiv f_0$) is a straight line  described by expressions
\begin{align}
\label{InnerGeodesic} t&=T+\tau,\\ \vec
r&=[z,x^1,x^2]=\left[f_0+\frac{\tau}{\sqrt{1+|\vec\nabla
 f|^2_0}}\ ,\ x^1_0-\frac{\tau
f_{x^1_0}}{\sqrt{1+|\vec\nabla f|^2_0}}\ ,\ x^2_0-\frac{\tau
f_{x^2_0}}{\sqrt{1+|\vec\nabla f|^2_0}}\right]\notag
\end{align}
in which subscript ``0" denotes values on the surface $\M_I$ and
$\tau$ is a time parameter.

Such geodesics transfer the trapped surface $\M_I$\ on the
distance $\tau$ to the surface $\M_I(\tau)$. Introducing a metric
tensor $g_{ij}$ on the  $\M_I(\tau)$ as
\begin{equation}\label{M_Imetric}
(\d\vec r,\d\vec r)=g_{ij}\d{}x_0^i\d{}x_0^j
\end{equation}
 we may write an area
of $\M_I(\tau)$
\begin{equation}\label{InnerArea}
S_I(\tau)= \int \sqrt{\left|g_{11}g_{22}-g_{12}g_{12}\right|}\
\d{}x^1_0\d{}x^2_0\ .
\end{equation}

Light geodesics that cross orthogonally a trapped surface converge
locally in the future-time direction \cite{Penrose:68book}. This
suggests that the area $S_I(\tau)$ decreases as the small distance
$|\tau|$ increases for both signs of $\tau$. So, an equation for
$f$ can be obtained by expanding the right part of
\eqref{InnerArea} as a power series in $\tau$ and setting a
$\tau$-linear term equal to zero \cite{Vasilenko:0305}
\begin{equation}
 \label{InnerTrapSurfaceEq}
\bigl(1+{f_{x^2}}^2\bigr)f_{x^1x^1}+\bigl(1+{f_{x^1}}^2\bigr)f_{x^2x^2}
 -2f_{x^1}f_{x^2}f_{x^1x^2}=0\ .
\end{equation}

In fact, this calculation is equivalent to finding a form of a
soap film or minimal surface and \eqref{InnerTrapSurfaceEq} is a
well known equation for a two-dimensional minimal surface in
three-dimensional euclidian space
\cite{Dubrovin/Novikov/Fomenko:1979book}.

To find a form of the inner trapped surface $\M_I$ it is necessary
to solve the equation \eqref{InnerTrapSurfaceEq} in the
region~$R_I$ with a function  $f$ taking values $\pm T$ on the
border contours $\C_{\pm}$. We shall search the function
$f(x^1,x^2)=z$ in an implicit form using the following ansatz
\begin{equation}\label{Ansatz}
\cosh\left[k\left(x^1-x_m(kz)\right)\right]=p(kz)\cos\left(kx^2\right)\
.
\end{equation}
It gives a solution of \eqref{InnerTrapSurfaceEq} if functions
$x_m$ and $p$ satisfy equations
\begin{equation}\label{AnsatzEquation}
{x_m}''\left(p^2-1\right)-2{x_m}'pp'=0\ ,\quad
\left(p''-p\left(1+{{x_m}'}^2\right)\right)\left(p^2-1\right)-2p{p'}^2=0\
.
\end{equation}
Solution of \eqref{AnsatzEquation} may be written in a parametric
form as
\begin{align}\label{Solution}
kz(p)&={\rm
sign}(z)\int^p_{p_0}\;\cfrac{p_0^2-1}{\xi^2-1}\;\cfrac{\d\xi}{\sqrt{(\xi^2-p_0^2)
\left(k^2c^2+\cfrac{p_0^2-1}{\xi^2-1}\right)}}\ ,\\
x_m(p)&=-{\rm sign}(z)\int^p_{p_0}\;\cfrac{c\,
\d\xi}{\sqrt{(\xi^2-p_0^2)
\left(k^2c^2+\cfrac{p_0^2-1}{\xi^2-1}\right)}}\ .
\end{align}
Parameters $c$, $p_0$ depend on $T$ and are determined by trapped
surface continuity conditions on the borders of the inner and
outer regions; that is, \eqref{Ansatz} must take the form of
\eqref{Contours-C} on $\C_{\pm}$
\begin{equation}\label{BorderConditions}
|z(p_b)|=-T\ ,\quad |x_m(p_b)|=x_b\ ,\quad p_b=e^{k(b/2-x_b)}\ .
\end{equation}

\begin{figure}[h]
\begin{center}
\includegraphics*[scale=0.7,clip]{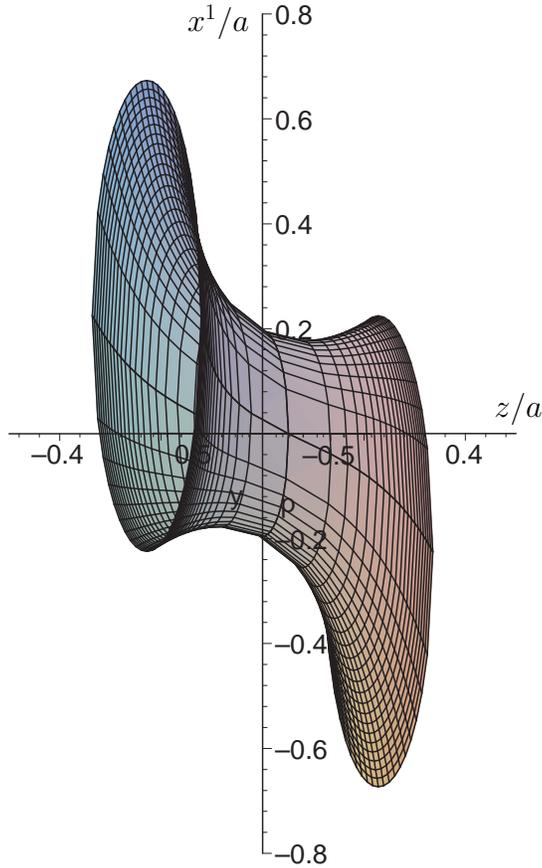}
\caption{\small The trapped surface for $b=b_{max}=0.8045a$ at the
moment of appearance $T=T_{min}(b_{max})=0.228a$\ .}
\label{fig:Surface3d}
\end{center}
\end{figure}

Fig.~\ref{fig:Surface3d} displays an example of the inner trapped
surface shape.

%


\section{Trapped surface formation}
\label{Trap surface formation}
\begin{figure}[h]
\begin{center}
\includegraphics*[scale=0.8,clip]{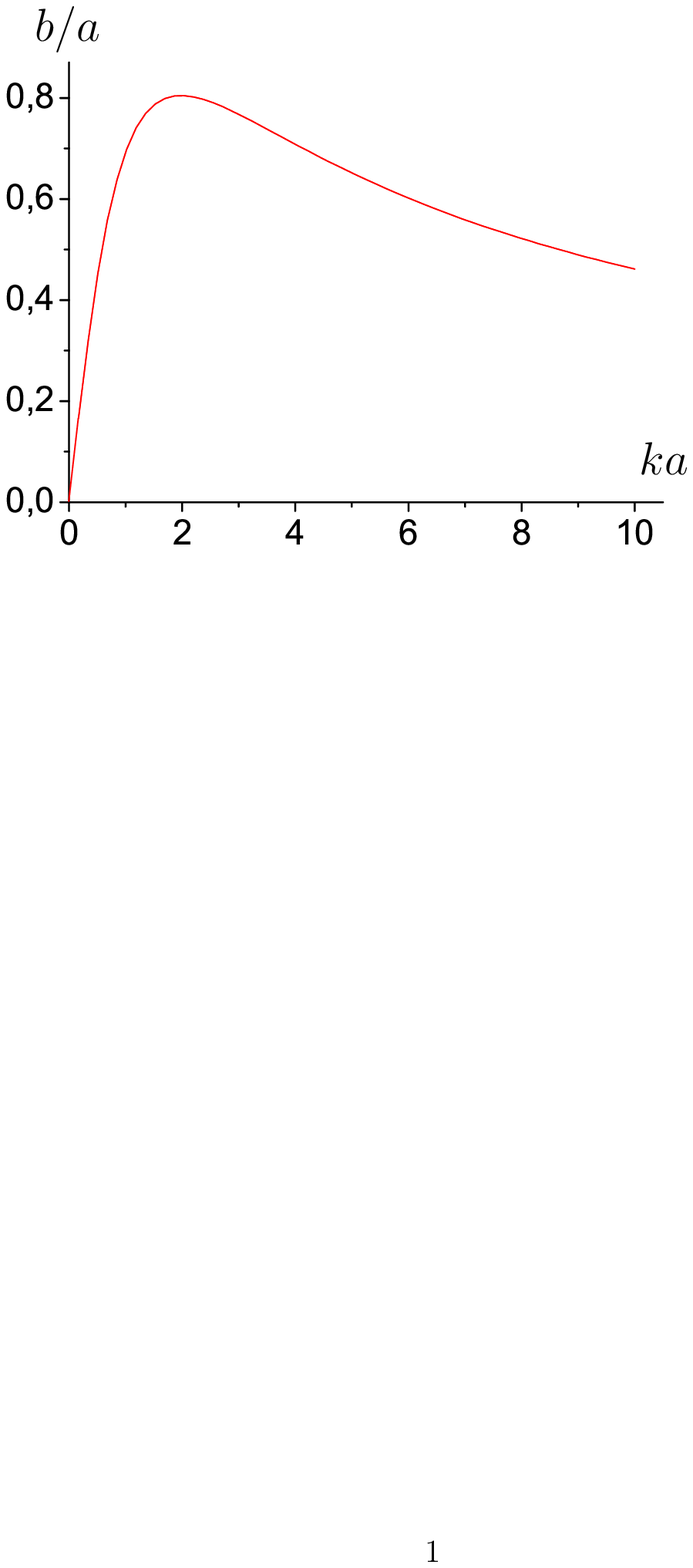}
\caption{\small The dependence of $b/a$ on $ka$.
$(b/a)_{max}=0.8047$ at $ka=1.9803$} \label{fig:bVsk}
\end{center}
\end{figure}

 The dependence \eqref{Function-k}
of $b/a$ on $ka$ (Fig.~\ref{fig:bVsk}) shows that for a given $b$
there are two values of $k: k_1(b)<k_2(b)$ . The impact parameter
$b$ reaches its maximum value $b_{max}=0.8047a$ at $k=1.98/a$.
This limits a cross-section of a black hole  production in a two
shocks impact \cite{Eardley/Giddings:0201}.

\begin{figure}[h]
\begin{center}
\includegraphics*{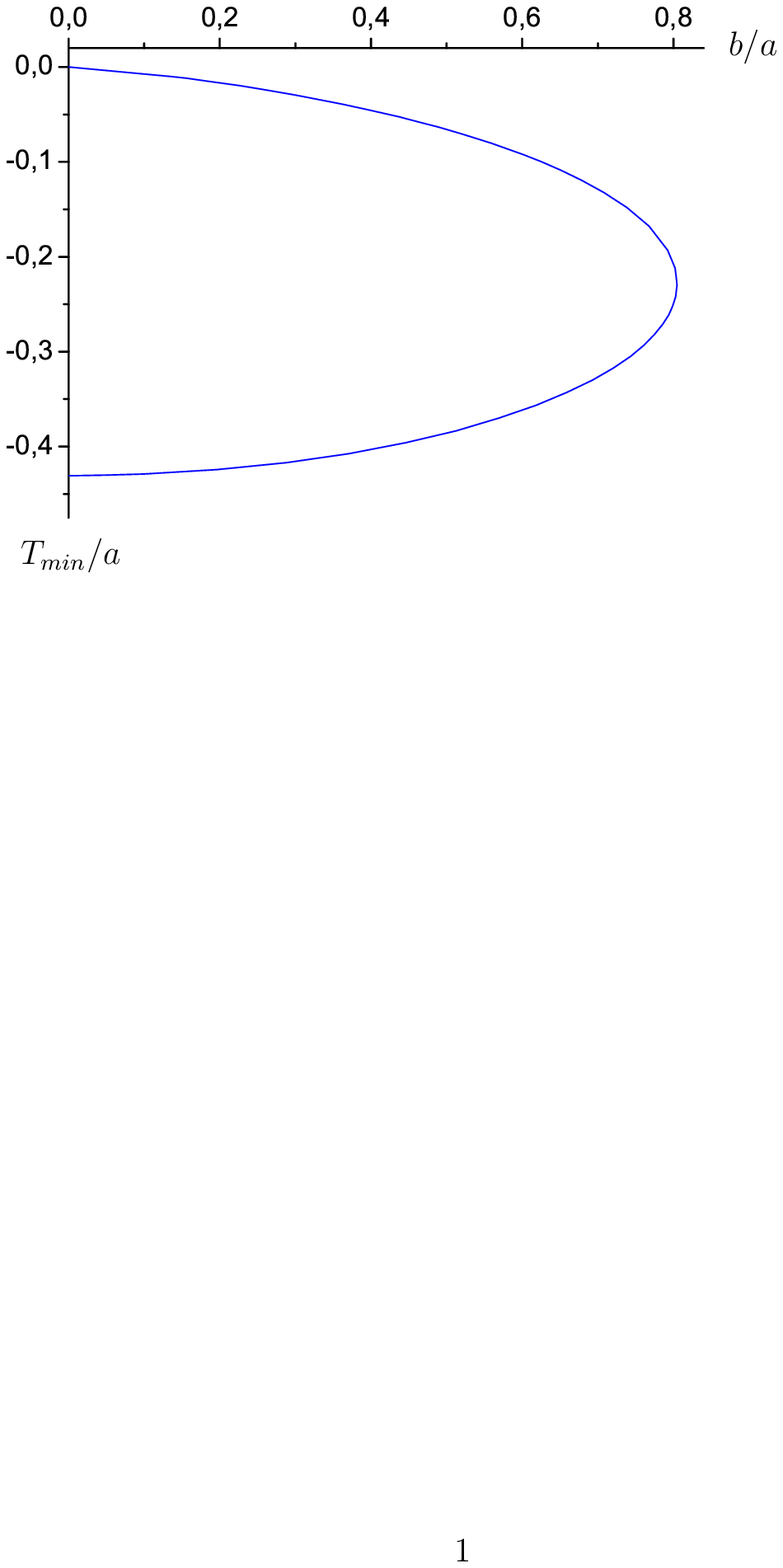}
\caption{\small The dependence of $T_{min}/a$ on $b/a$ .}
\label{fig:TminVSb}
\end{center}
\end{figure}

\begin{figure}[h]
\begin{center}
\includegraphics*{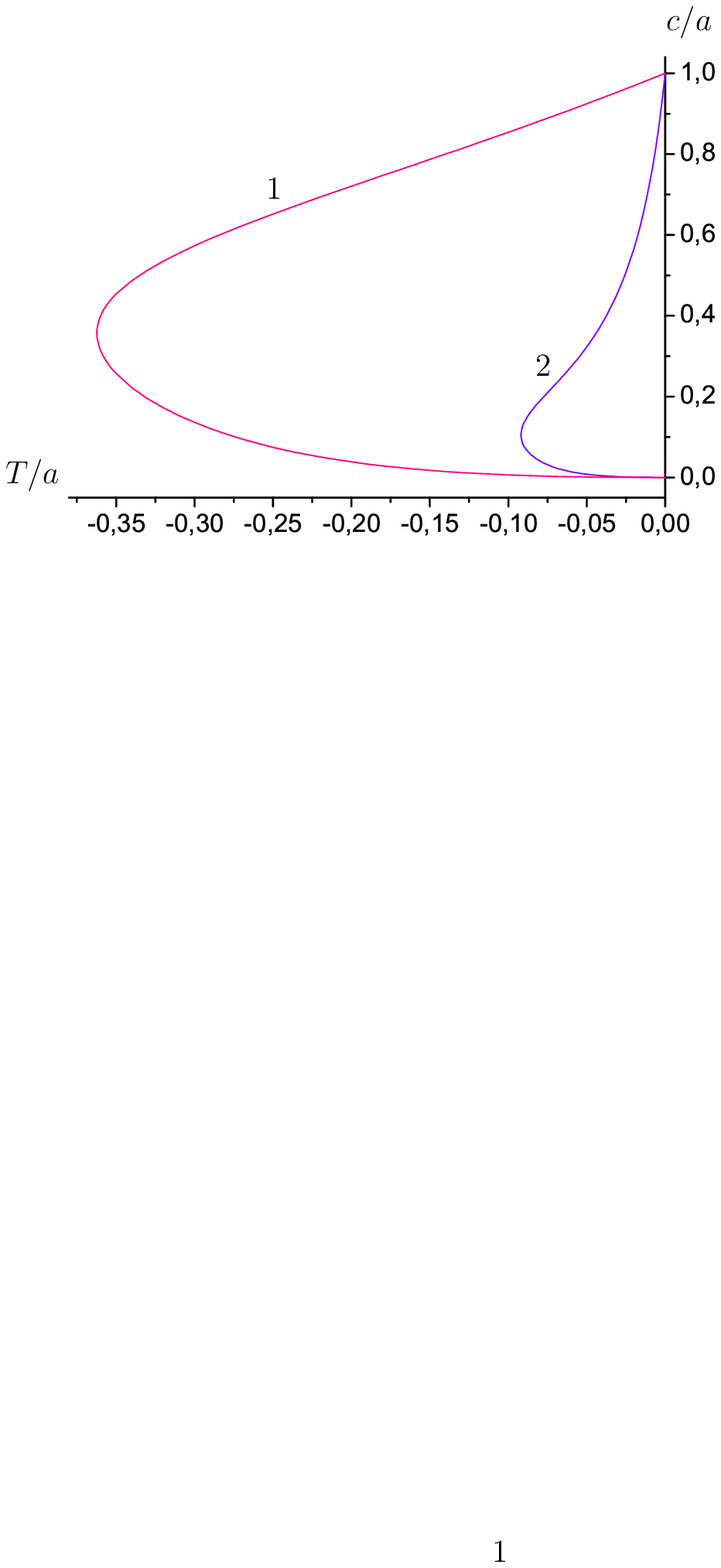}
\caption{\small The dependences of $c/a$ on $T/a$ for $b/a=0.6$. 1
-- $k_1=0.76$, 2 -- $k_2=6.04$ .} \label{fig:cVsT}
\end{center}
\end{figure}

The trapped surface appears at $T=T_{min}$. For a given $b$ there
are two values of $T_{min}: T_{min\,1}<T_{min\,2}$ corresponding
to $k_1(b)$ and $k_2(b)$ respectively (Fig.~\ref{fig:TminVSb}).
The relation between parameter $c$ and $T$ is also two-valued
(Fig.~\ref{fig:cVsT}).

\begin{figure}[h]
\begin{center}
\includegraphics*[scale=0.7,clip]{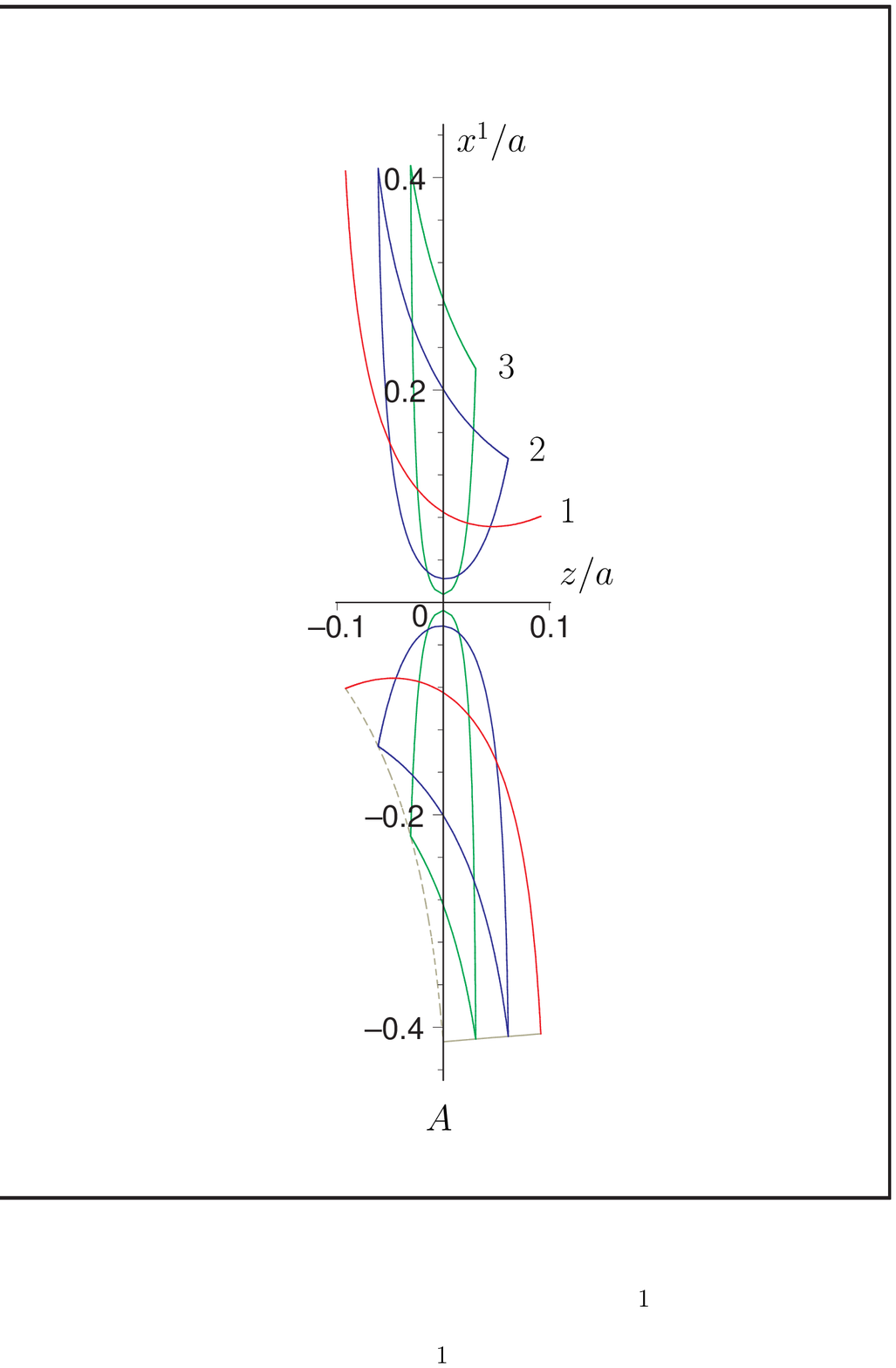}
\includegraphics*[scale=0.7,clip]{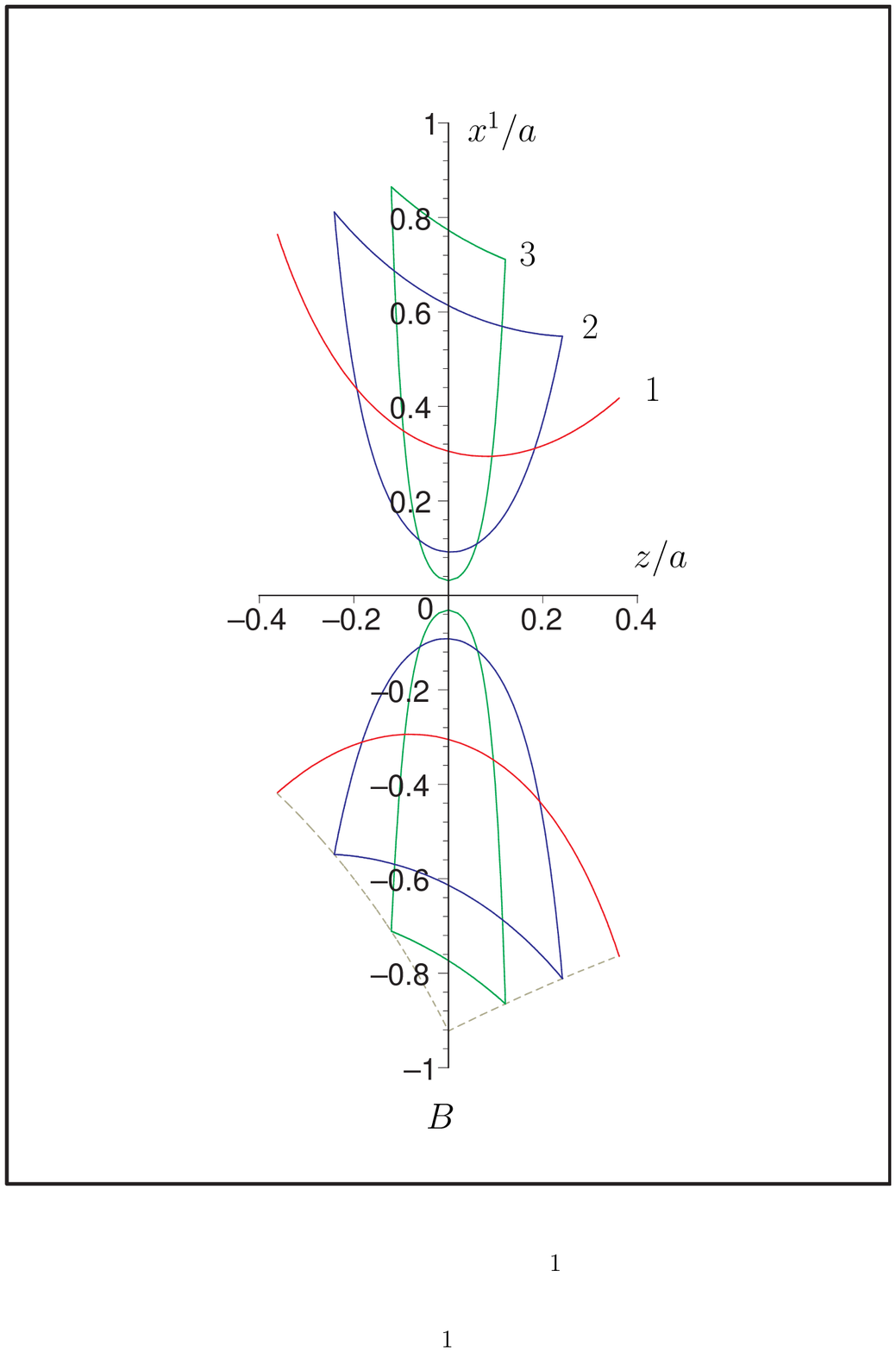}
\caption{\small The ($x^2=0$)-sections of the  trapped surfaces
for $b=0.6$ at different times $T$. $A$ --- $k_2=6.04,\
T_{min\,2}/a=-0.092$; $B$ --- $k_1=0.76,\ T_{min\,2}/a=-0.362$\ .
1 --- $T=T_{min}$; 2 --- $T=2T_{min}/3$; 3 --- $T=T_{min}/3$. For
$T>T_{min}$ there are external and internal surfaces corresponding
to $c>c(T_{min)}$ and $c<c(T_{min})$ respectively\
(Fig.\ref{fig:cVsT}). In the low parts of the graphics
trajectories of endpoints are shown by dot lines.}
\label{fig:Section}
\end{center}
\end{figure}

On the whole, the process of the horizon formation looks as
follows. The horizon appears at $T=T_{min\,1}$. At
$T_{min\,1}<T<T_{min\,2}$ two trapped surfaces exist. One on them
(internal) is located inside another one (external). Their
evolution is illustrated in Fig.~\ref{fig:Section}B, where
($x^2=0$)-sections of the trapped surfaces at different times are
presented. As time $T$ increases ($z=0$)-diameter of the internal
and external surfaces decreases and increases respectively. At
$T=T_{min\,2}$ two other trapped surfaces appear with similar
properties (Fig.~\ref{fig:Section}A).

So, at $T>T_{min\,2}$ we have four different trap surfaces. The
biggest one of them is the external surface that appears at
$T=T_{min\,1}$.  It may be considered as a horizon. Others are
spaced inside it. It is possible that in more realistic model the
number of the trapped surfaces will be less and the horizon may be
defined more precisely.

According to the results of \cite{Vasilenko:0305} where the
head-on collision was investigated we may expect that the results
obtained above may be qualitatively correct in higher dimensions.

%

%

%

%

\end{document}